\documentclass[aps,prd,reprint,showpacs,longbibliography]{revtex4-2}
\usepackage{amsmath,amssymb,amsthm,graphicx}
\usepackage[mathscr]{eucal}
\usepackage[colorlinks,linkcolor=blue,
anchorcolor=blue,citecolor=green]{hyperref}
\usepackage[dvipsnames,usenames]{color}
\newcommand{\no}{\nonumber}
\newcommand{\be}{\begin{equation}}
	\newcommand{\ee}{\end{equation}}
\newcommand{\ba}{\begin{eqnarray}}
	\newcommand{\ea}{\end{eqnarray}}
\usepackage{epsf,epsfig,graphics,soul}
\usepackage{verbatim,color,ulem}

\begin{document}
\title{Analogous Hawking radiation from gapped excitations in a transonic flow of  binary Bose-Einstein condensates}
\date{\today}
\author{Wei-Can Syu}
\email{syuweican@gmail.com}
\affiliation{Department of
	Physics, National Dong-Hwa University, Hualien 974301, Taiwan, R.O.C.}
\author{Da-Shin Lee}
\email{dslee@gms.ndhu.edu.tw}
\affiliation{Department of
	Physics, National Dong-Hwa University, Hualien 974301, Taiwan, R.O.C.}

\begin{abstract}
We have studied analytically the approximate  solutions to the gapped mode equations in the hydrodynamic regime for a class of binary Bose-Einstein condensate acoustic black holes. The horizon from the transonic flow is formed by manipulating the phonon sound speed and the flow velocity with the experimentally accessible parameters.  The asymptotic modes of various scattering processes are constructed  from which to obtain  scattering coefficients and then to further decompose the field operator in terms of the asymptotic states. Also,  the Unruh state
is introduced to be the appropriate state for the description of
gravitational collapse of the black hole.
The particle densities of the outgoing modes are computed. The effective energy gap term in the dispersion relation of the gapped excitations introduces the
threshold frequency $\omega_r$ in the subsonic regime, below which the propagating modes do not exist.
Thus, the particle spectrum of the analogous Hawking modes in the exterior of the horizon of the subsonic region significantly deviates from that of the gapless
cases near the threshold frequency due to the modified graybody factor, which vanishes as the
mode frequency is below  $\omega_r$.
 However, in the interior region of the  horizon of the supersonic region, the spectrum of  the  particle production of the Hawking partner has the nonthermal feature. The correlators between the analog Hawking mode and its partner of relevance to the experimental observations are also investigated and show some peaks near the
threshold frequency $\omega_r$ resulting from  the gap energy term  to be seen in future experiments.
\end{abstract}

\maketitle
\newpage
\section{INTRODUCTION}\label{sec_introduction}
The evaporation of black holes has been predicted by Hawking through the emission of a thermal flux of radiation, thereby reducing its mass \cite{Hawking:1975aa}. However, the expected Hawking temperature is of order of $T_H=6.0 \times 10^{-8} \text{K}$ for an astrophysical black hole, which is several orders of magnitude smaller than the cosmic microwave background temperature $T_{cmb} \approx 3 \text{K}$. Thus, the detection of Hawking radiation in an astrophysical context is extremely unlikely.
The program of the analog models of gravity due to the pioneering work of Unruh is an attempt to implement laboratory systems to mimic various phenomena that happen  in the interplay between general relativity and quantum field theory such as  in black holes and the early Universe.
The aim is for devising experiments of  real laboratory tests that provide insights in phenomena and further probe the structure of curved-space quantum field theory.
 From a prospective of the analog gravity, Unruh realized that sound waves in a moving fluid can be analogous to light waves in curved spacetime where the supersonic fluid can generate acoustic black holes with acoustic horizons \cite{Unruh1981}.
Thus, the existence of analogous photonic Hawking radiation can be theoretically demonstrated.  The work of Ref.~\cite{Jeff2019} is a first experimental observation of Hawking radiation extracted from correlations of the collective excitations that agree with a thermal spectrum with the temperature estimated from  analog surface gravity.
Also, the time dependence of the Hawking radiation in an analog black hole is observed in Ref.~\cite{Kolobov:2021wd}.
With the advent of the experimental study of  binary Bose-Einstein condensates (BECs) in Refs.~\cite{Kim2020,Cominotti2022}, the condensates of cold atoms at zero temperature in such  tunable  systems  have been explored with  the Rabi transition between  atomic hyperfine states where the system can be represented by a coupled two-field model of gapless excitations and gapped excitations. The binary BECs have been adopted as an analog model to theoretically  mimic quantum phenomena in the early Universe and/or black holes \cite{Visser2005,Fischer2004,Liberati2006,syu2022}.
In our previous work \cite{syu2022} we set up the configuration of the supersonic and subsonic regimes with constant flow velocity where the acoustic horizon is established between them in the elongated two-component BECs, trying to stimulate analogous Hawking radiation, in particular due to the gapped excitations. In this work, we consider the same model where  the sound speed and flow velocity profiles are spatial dependent to generate the acoustic horizon with reference to the experiments in Refs.~\cite{Jeff2016,Jeff2019} and the theoretical studies in Ref.~\cite{Fabbri2016,Antonin2018}.   Here we briefly review in what conditions of the  coupling constants and condensate wave functions of the spatial dependence  two collective excitations can be decoupled
by following the work of  \cite{Visser2005}.
We  then consider the dispersion relation of the gapped modes with the $k^2$ term in a  very long wavelength approximation that behaves relativistically.  Some {profiles}  of the spatial-dependent sound speed and the flow velocity are  specified where the approximate analytic solutions of  the mode equations of the gapped excitations can be treatable. The Unruh state
is introduced as the appropriate state for the description of
gravitational collapse of the black hole.
The particle spectrum of the outgoing modes and their correlators are computed from which to
 further discuss  the relevance to the experimental measurements.

We organize this paper as follows.  In Sec.~\ref{ABH2}, we introduce the model of a binary BEC system   and give a review of the approach to decouple the gapless and gapped excitations.
The particular  sound speed  and flow velocity profiles later are introduced to establish the acoustic horizon of the black hole. In addition, various asymptotic modes are identified.
In Sec.~\ref{ABH3},  the approximate analytical solution to the mode equations of the gapped excitations of interest is found from which the  reflection and transmission coefficients for each scattering process are obtained. Also,  the Unruh state
is introduced.
In Sec.~\ref{ABH4},  the field of the gapped excitations is expanded in terms of either incoming or outgoing modes whereas the  corresponding $S$-matrix and the Bogoliubov transformations of the creation and annihilation operators between the incoming and outgoing modes are constructed.
Section~\ref{ABH5} is devoted to obtaining the particle density of outgoing modes and their correlators.
We conclude the work in Sec.~\ref{ABH6}.

\section{{Review of Mode decoupling in TWO-component BECs}}\label{ABH2}
The effective  mass term  of the quantum field theory can be considered as  the gapped energy term of the gapped excitations in a binary BECs system in the hydrodynamic approximation\cite{Visser2005,Fischer2004, syu2022}. Here we consider the binary BEC of the same atoms in two different internal hyperfine states.  This class of the two-component BEC systems with the Rabi interaction exhibits two types of  excitation on condensates: the gapless excitation due to the ``in-phase'' oscillations between  two respective density waves and  the gapped excitation stemming from the ``out-of-phase'' oscillations of the density waves with additional the Rabi transition, which are respectively analogous of the Goldstone modes and the Higgs modes in particle physics.  Here we briefly review how two excitations are decoupled under certain conditions of the spatial-dependent
condensate wave functions and the coupling constants  \cite{Visser2005}.
With the unit $\hbar=k_\text{B}=1$ throughout this paper, the  time-dependent equations of motion in $1+3$ dimensions are expressed by
\begin{subequations}\label{GP}
\begin{align}	i\partial_t\hat{\Psi}_1=&\left[-\frac{1}{2m}\vec\nabla^2+V_1( \vec x)+g_{11}\hat{\Psi}_1^\dagger\hat{\Psi}_1+g_{12}\hat{\Psi}_2^\dagger\hat{\Psi}_2\right]\hat{\Psi}_1\no\\
	&-\frac{\Omega}{2}\hat{\Psi}_2,\\
i\partial_t\hat{\Psi}_2=&\left[-\frac{1}{2m}\vec\nabla^2+V_2 ( \vec x)+g_{22}\hat{\Psi}_2^\dagger\hat{\Psi}_2+g_{12}\hat{\Psi}_1^\dagger\hat{\Psi}_1\right]\hat{\Psi}_2\no\\
&-\frac{\Omega}{2}\hat{\Psi}_1,
\end{align}
where $m$ is atomic mass, and $V_1,\,V_2$ are the external potentials  on the hyperfine states $1$ and $2$, respectively.
Additionally, $g_{11},\, g_{22}$, and $g_{12}$ are the interaction strengths of atoms between the same hyperfine states and different hyperfine states, respectively. The coupling strengths  are related with the scattering lengths. Experimentally, the values of  scattering lengths can be tuned using Feshbach resonances such as two hyperfine states of $^{87}\text{Rb}$~\cite{Myatt1997,Hall1998,Papp2008,Tojo2010}. We also introduce a Rabi coupling term by shining the laser field or applying the radio wave with the strength given by the Rabi frequency $\Omega$ \cite{Matthews1998,Nicklas2011}.
\end{subequations}
The condensate wave functions are given by the expectation value of the field operator $\langle\hat{\Psi}_i \rangle$
\begin{align}
	\langle \hat{\Psi}_i\rangle=\sqrt{\rho_{i}}\, e^{ i\theta_{i}-i \mu t}
	\label{perturb}
\end{align}
 with the chemical potential $\mu$. The condensate flow velocities are given by $\vec \nabla \theta_{i} (x)/m=\vec v_i (x)$  ($i=1,2$).
 The equations for $\rho_i$ and $\theta_i$ of the condensate wave functions can be found in Refs.~\cite{Visser2005,Liberati2006,syu2022}.
 The perturbations around the stationary wave function are defined through
\begin{align}
\hat{\Psi}_i=\langle\hat{\Psi}_i \rangle(1+\hat{\phi}_i) \, ,
\label{psi}
\end{align}
{
 where 	the fluctuation fields in \eqref{psi} can be decomposed  in terms of the density and the phase  as
\begin{align}
 		\hat{\phi}_i=\delta{\hat{n}_i}+i\delta\hat{\theta}_i=\frac{\delta\hat{\rho}_i}{2\rho_{i}}+i\delta\hat{\theta}_i\,.\label{phi}
 	\end{align}
 {Substitutions of \eqref{psi} and \eqref{phi} into \eqref{GP} }  give the coupled equations of two states 1 and 2.
For the general spatial-dependent condensate wave functions as well as the coupling strengths,
   it is found that the above equations can be decoupled  by choosing  $\rho_{1}=\rho_2=\rho$, $\theta_{1}=\theta_2=\theta$, and
$g_{11}=g_{22}=g$ \cite{Visser2005}. The  chosen scattering parameters in the binary systems can have a miscible state of background condensates \cite{Hamner2011,Hamner2013}.
The detailed analysis of the choice of the parameters can be found in our previous work \cite{Syu2019}.
One can therefore define
\begin{subequations}
\begin{align}
	&\delta \hat n_{d/p}=\frac{1}{\sqrt{2}}\left(\delta \hat{n}_1\pm \delta \hat{n}_2\right),\\
	&\delta \hat {\theta}_{d/p}=\frac{1}{\sqrt{2}}\left(\delta \hat{\theta}_1\pm \delta \hat{\theta}_2\right),
\end{align}
\end{subequations}
where the subscript $d$ ($p$)  refers to the density (polarization) fluctuations. The decoupled equations are shown to be
\begin{subequations}\label{n_gapless}
\begin{align}
	&	\partial_t\delta\hat{\theta}_d=\frac{1}{2m\rho}\vec\nabla\cdot( \rho\vec \nabla \delta \hat{n}_d)-\vec{v}\cdot \vec\nabla \delta\hat{\theta}_d-2(g+g_{12})\rho\delta\hat{n}_d, \label{theta_gapless}\\
	&	\partial_t\delta\hat{n}_d=-\frac{1}{2m\rho}\vec\nabla\cdot( \rho\vec \nabla \delta \hat{\theta}_d)-\vec{v}\cdot\vec\nabla \delta\hat{n}_d,
\end{align}
\end{subequations}
and
\begin{subequations}\label{gapped}
	\begin{align}
		\partial_t\delta\hat{\theta}_p=&\frac{1}{2m\rho}\vec\nabla\cdot( \rho\vec \nabla \delta \hat{n}_p)-\vec{v}\cdot \vec\nabla \delta\hat{\theta}_p-[2(g-g_{12})\rho\no\\
		&+\Omega]\delta\hat n_p,\label{theta_gapped}\\
		\partial_t\delta\hat{n}_p=&-\frac{1}{2m\rho}\vec\nabla\cdot( \rho\vec \nabla \delta \hat{\theta}_p)-\vec{v}\cdot\vec\nabla \delta\hat{n}_p+\Omega\delta\hat\theta_p.\label{n_gapped}
\end{align}
\end{subequations}
The analog Hawking radiation arising from the gapless modes  has been studied in the literature \cite{Balbinot2008,Recati2009,Macher2009, Larre2012}.
Here we mainly focus on the gapped modes given by the polarization excitations in \eqref{gapped} where from now on the subscript $p$ is dropped out for simplifying the notation.
 In this paper, we consider the transonic flow, which is accelerated by manipulating the condensate density $\rho(x)$ with the spatial dependence obtained from a sharp external potential \cite{Jeff2016, Jeff2019} and spatial-dependent interaction strength $g_{12}(x)$.  The experimentally spatial variation of the interaction strengths is challenging but feasible \cite{Clark2015,Arunkumar2019,Carli2020}.}

One thus combines \eqref{theta_gapped} and \eqref{n_gapped} to obtain
 \begin{align}
-\left(\partial_t +\vec\nabla \vec v\right)\frac{\rho}{mc^2}\bigg(\partial_t +\vec{v}\,\vec\nabla\bigg)\delta\hat{\theta}+\vec\nabla \frac{\rho}{m}\vec\nabla\delta\hat{\theta}-2\rho \Omega\delta\hat{\theta}=0
\label{bogo_gennes}
\end{align}
with the spatial-dependent sound speed  $c(x)=\sqrt{[(g-g_{12})\rho(x)+\Omega]/m}$.
To further express \eqref{bogo_gennes} as the form of  the Klein-Gordon  equation,
 the equation can be rewritten as
\begin{align}
	&\bigg(\square -\frac{mm_\text{eff}^2}{\rho c}\bigg)\delta\hat{\theta}=\frac{1}{\sqrt{-\mathbf{g}}}\partial_\mu\left(\sqrt{-\mathbf{g}}\,\mathbf{g}^{\mu\nu}\partial_\nu \delta\hat{\theta}\right)-\frac{mm_\text{eff}^2}{\rho c}\delta\hat{\theta}\no\\&=0
	\label{KGE}
\end{align}
with the gapped energy $ m_\text{eff}(x)=\sqrt{2(g-g_{12})\rho(x)\Omega+\Omega^2}$ $= \sqrt{2m c^2(x) \Omega} $.
The acoustic metric is
\begin{align}\label{metric}
	ds^2=\frac{\rho }{mc}\left[-(c^2-v^2)dt^2-2\vec{v}\cdot d\vec{x}dt+d{x}^2+dy^2+dz^2\right],
\end{align}
 where we choose  the direction of the flow along the $x$ direction,  $\vec{v}(x)=-v(x)\hat{x}\,\,[v(x)>0]$.
It is then assumed that the system  can be treated in the pseudo-one-dimension by applying a strong cigar-shape trap potential where the size of the trap $L_x$ along the axial direction, say in the $x$ direction is much larger than the size of $L_r$ along the radial direction \cite{Hamner2011, Hamner2013, Cominotti2022}.
Later, we will choose  the profile of the sound speed and the flow velocity so that the Klein-Gordon equation can be  treatable  analytically in some approximations to be discussed later \cite{Fabbri2016,Roberto2019,Dudley2018}.
Using the transformation to define the time $\tau$ from the laboratory time $t$ as
\begin{align}
	\tau= t - \int dx \frac{v}{c^2-v^2}
\end{align}
to rewrite the metric \eqref{metric} restricted in one dimension along the $x$ direction as
\begin{align}
	ds^2=\frac{\rho }{mc}[-({c^2-v^2})d\tau^2+\frac{c^2}{c^2-v^2}dx^2],
\end{align}
the corresponding Klein-Gordon equation becomes

	\begin{align}
		\left[	\frac{-c}{\rho\left(c^2-v^2\right)}\partial_{\tau}^2 +\partial_x\left(\frac{c^2-v^2}{\rho\,c}\partial_x\right)-\frac{m_\text{eff}^2}{\rho\,c}\right] \delta\hat\theta(x,\tau)	=0.\label{KGE2}
\end{align}
According to Ref.~\cite{Fabbri2016}, we perform  the further variable transformation
\begin{align}
dx =\frac{1}{\rho}\left(1-\frac{v^2}{c^2}\right)dz
\label{dxtodz}
\end{align}
giving  the  metric

\begin{align}\label{metric_conformal}
	ds^2=\frac{\rho }{mc} (c^2-v^2)\left[-d\tau^2+\frac{1}{\rho^2 c^2}{dz^2}\right].
\end{align}
We further use the continuity equation of the gapped excitations  $v\rho=\text{constant}$ resulting from the respective continuity equations for states $1$ and $2$  with equal phases and densities between them  to set  $\rho=1/v$  and rewrite \eqref{KGE2} to be
\begin{align}
	\left[\frac{v^2}{c^2}\partial_\tau^2-\partial^2_z+v^2\left(1-\frac{v^2}{c^2}\right)2m\Omega\right]\delta\hat\theta(z,\tau)=0.
	\label{mode_eq}
\end{align}
The third term  of \eqref{mode_eq} is induced by the gap energy, giving the analog mass term of the relativistic quantum scalar field.
 The  mass term is positive (negative) in the subsonic (supersonic) region, which implies that there will exist the threshold frequency
in the subsonic region beyond which to have the propagating modes \cite{Antonin2012,Jannes2011}.

\begin{figure}[t]
	\includegraphics[width=\columnwidth]{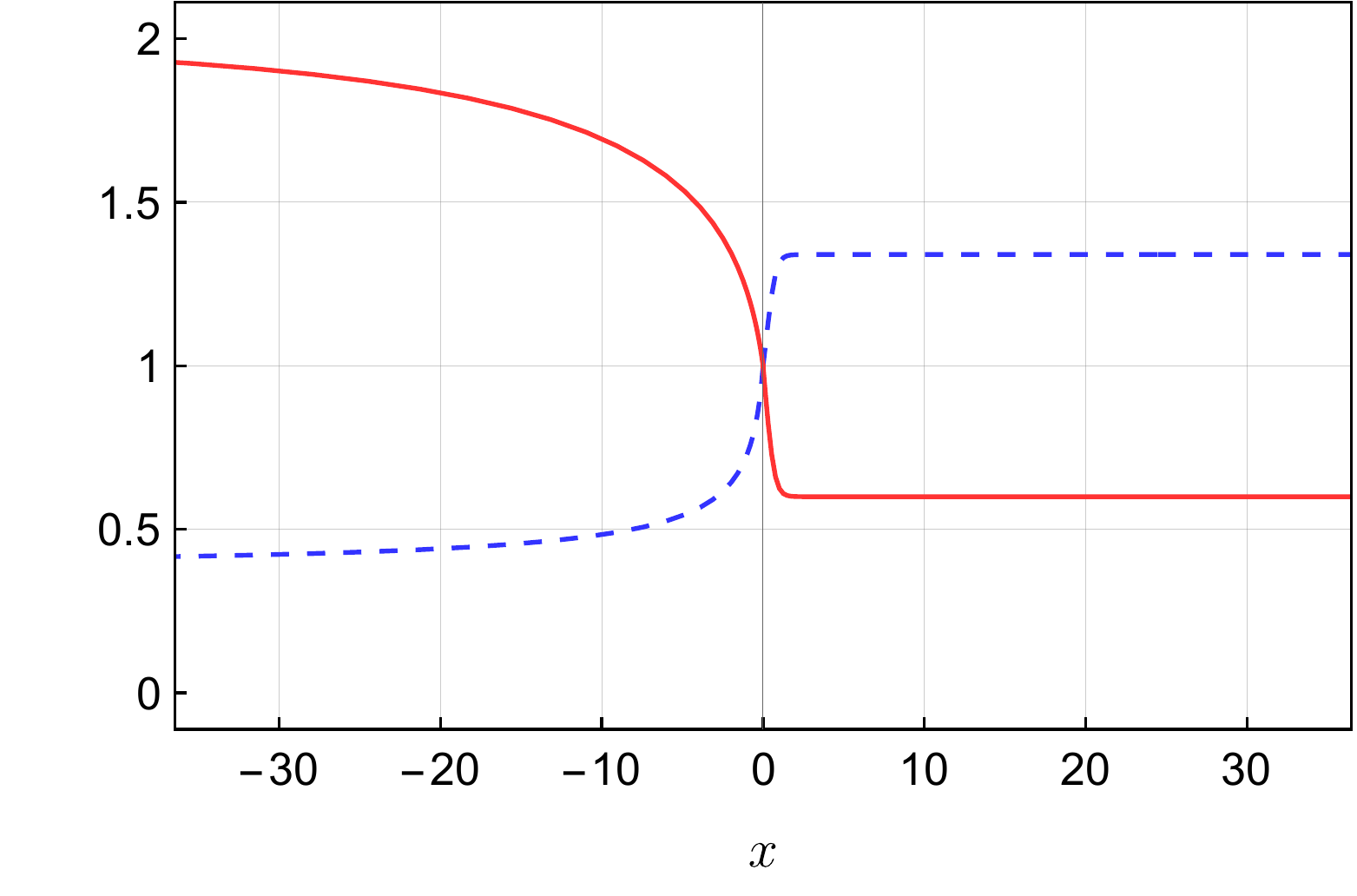}
	\caption{Profiles of the sound speed $c(x)$ (blue dashed line)  and the flow velocity $v(x)$ (solid red line) are depicted from \eqref{cprofile_sup} and \eqref{cprofile_sub} with parameters $c_l=0.4,\,v_l=2,\,c_r=1.34,\,v_r=0.6$ with the reference value $v_0=1$ at the horizon. The Mach numbers are chosen to be $M_l=5$ and $M_r=0.44$ where $M_r=1/\sqrt{M_l}$ with reference to Refs.~\cite{Larre2012,Michel2016,Antonin2018}.}
	\label{fig_c}
	\includegraphics[width=\columnwidth]{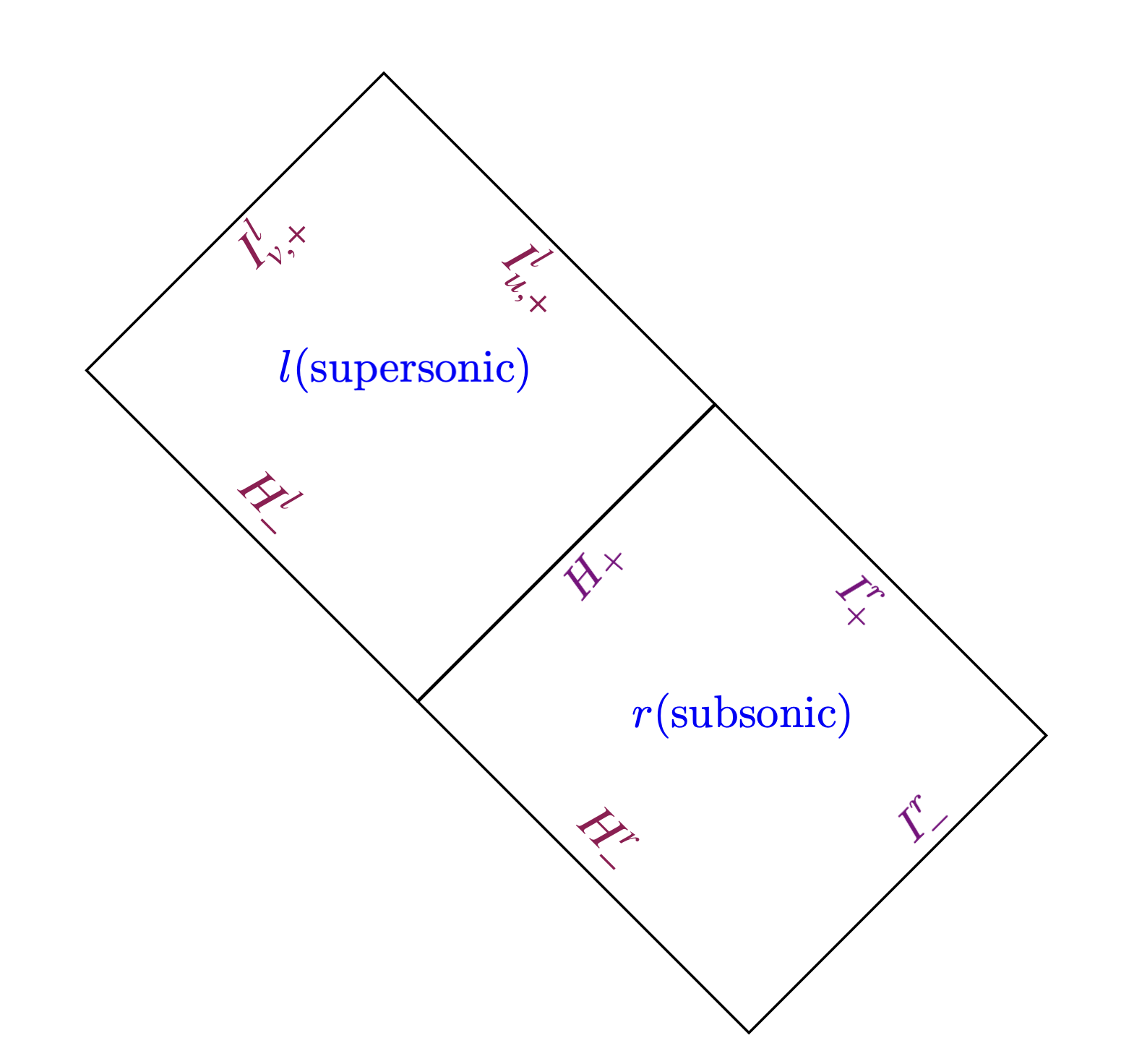}
	\caption{Schematic plot of the Penrose diagram.}
	\label{fig_penrose}
\end{figure}

 In the supersonic region ($v>c$), we assume that the sound speed and the flow velocity have the forms \cite{Fabbri2016,Antonin2018}
\begin{subequations}		\label{cprofile_sup}\begin{align}
		&c_\text{sup}(z)=\left[\frac{c_l^{-2}+v_0^{-2}}{2}+\frac{\left(c_l^{-2}-v_0^{-2}\right)}{2}\tanh{(\kappa z)}\right]^{-1/2},\\
		&	v_\text{sup}(z)=\left[\frac{v_l^{2}+v_0^{2}}{2}+\frac{\left(v_l^{2}-v_0^{2}\right)}{2}\tanh{(\kappa z)}\right]^{1/2}
\end{align}
\end{subequations}
with the asymptotic behaviors  $(v_\text{sup},\,c_\text{sup})_{z\rightarrow\infty}\rightarrow( v_l,\,c_l)$ for $x \rightarrow -\infty$ and $(v_\text{sup},\,c_\text{sup})_{z\rightarrow-\infty}\rightarrow(v_0,\,v_0)$ for $x\rightarrow 0$ of the horizon given from the spatial coordinate transformation (\ref{dxtodz}). In the subsonic region ($c>v$), they are assumed to be
\begin{subequations}\label{cprofile_sub}\begin{align}
		&c_\text{sub}(z)=\left[\frac{c_r^{-2}+v_0^{-2}}{2}+\frac{\left(c_r^{-2}-v_0^{-2}\right)}{2}\tanh{(\kappa z)}\right]^{-1/2},\\
		&	v_\text{sub}(z)=\left[\frac{v_r^{2}+v_0^{2}}{2}+\frac{\left(v_r^{2}-v_0^{2}\right)}{2}\tanh{(\kappa z)}\right]^{1/2},
\end{align}\end{subequations}
where $(v_\text{sub},c_\text{sub})_{z\rightarrow\infty}\rightarrow (v_r,c_r)$ for $ x\rightarrow +\infty$ and $(v_\text{sub},c_\text{sub})_{z\rightarrow-\infty}\rightarrow v_0$  for $ x\rightarrow 0$ of the horizon again from (\ref{dxtodz}).
The horizon from the transonic flow with the above profile  is formed by manipulating the phonon sound speed and the flow velocity with the experimentally accessible parameters  \cite{Jeff2016,Jeff2019}.
In Fig.~\ref{fig_c}, we show the transonic transition according to \eqref{cprofile_sup} and \eqref{cprofile_sub} under the transformation \eqref{dxtodz}.
Thus, in the asymptotical regions, the velocities reach respective constants where  the metric (\ref{metric_conformal}) is conformal to that of the Minkowski spacetime with the spatial coordinate in terms of  the rescaled $vdz/c$. One can define the incoming and outgoing states in these asymptotic regions of the Penrose diagram shown  in Ref.~\cite{Fabbri2016} and  Fig.~\ref{fig_penrose} in this paper.  The location of the analog horizon is at $x=0$   with the surface gravity
	\begin{align}
		\kappa=\frac{d(c-v)}{dx}\Big\vert_{x=0}
		\label{sur_gra}
\end{align}
that can be justified by substituting \eqref{dxtodz}, \eqref{cprofile_sup}, and \eqref{cprofile_sub} into \eqref{sur_gra}.

With the  profiles of a transonic flow \eqref{cprofile_sup} and \eqref{cprofile_sub}, the solutions of \eqref{mode_eq} in the subsonic and supersonic regions separately can be cast in a form of $\delta\theta (\tau,z)=e^{-i \omega \tau} \varphi_\omega (z)$, to therefore rewrite \eqref{mode_eq} as
\begin{align}
	\omega^2M_\text{a}^2(z)\eta_\text{a}^2(z)\varphi_\omega^\text{a}(z)+\frac{d^2\varphi_\omega^\text{a}(z)}{dz^2}=0\,,
	\label{mode_eq2}
\end{align}
where the index ``a'' refers to ``sup'' or ``sub'' for different regions. The quantity $\eta_\text{a} (z)$ is defined as
\begin{align}
	\eta_\text{a}(z)=\sqrt{1-(\pm)_\text{a} \frac{\omega_\text{a}^2(z)}{\omega^2}}
	\label{eta}
\end{align} with
\begin{equation}\label{omega_a}
\omega_\text{a}^2(z)\equiv2m\vert c_a^2(z)-v_a^2(z)\vert\,\Omega\,,
\end{equation} where $+$ ($-$) is  for the subsonic (supersonic) region.
We also express \eqref{mode_eq2} in terms of the Mach number defined by
\begin{align}
	M_\text{a}(z)=\frac{v_\text{a}(z)}{c_\text{a}(z)}.
\end{align}

\begin{figure}[b]
	\includegraphics[width=\columnwidth]{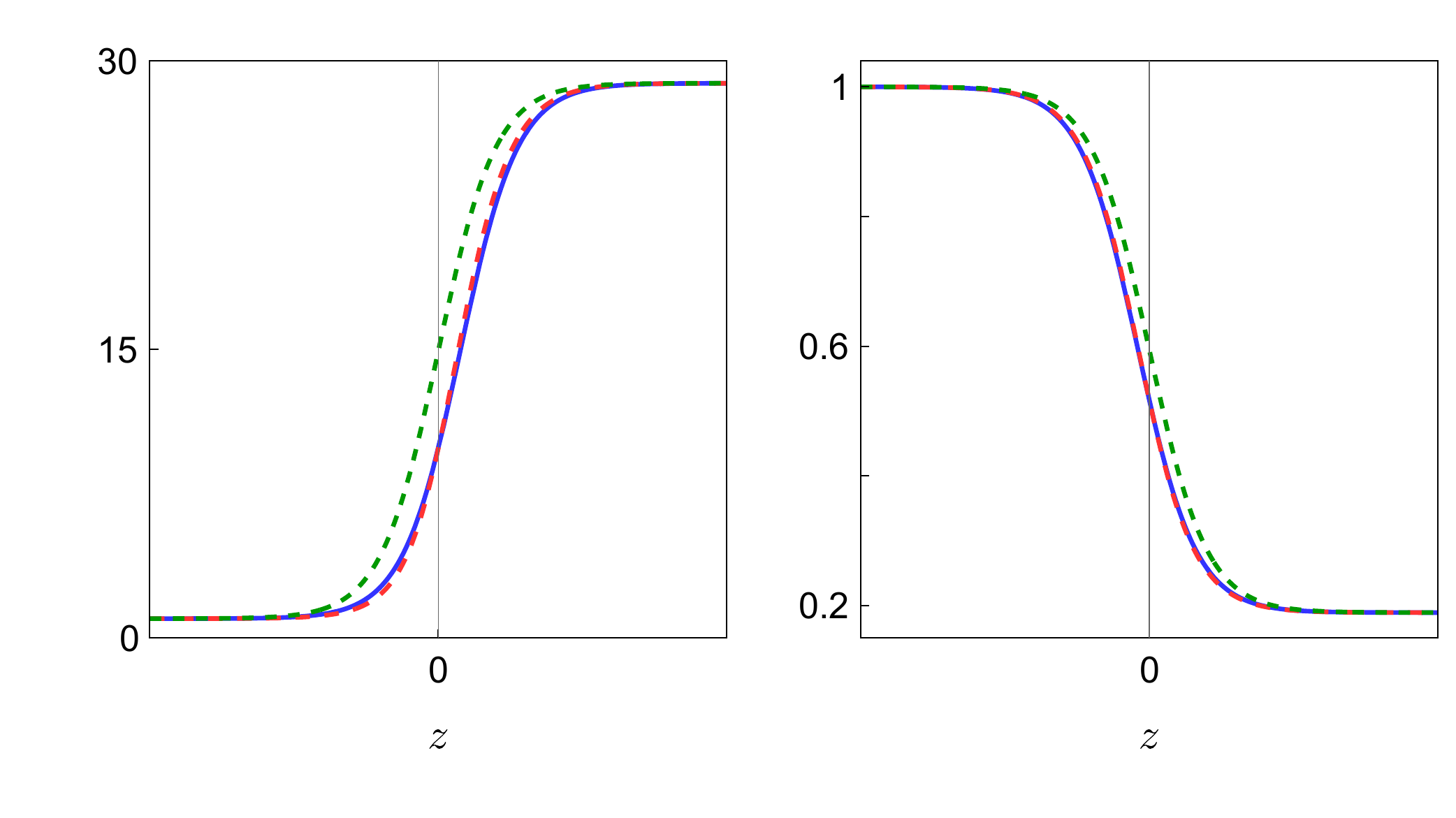}
	\caption{{Comparison of the exact $M^2\eta^2(z)$ (blue line) in \eqref{mode_eq2} computed from the sound speed and the flow velocity in  \eqref{cprofile_sup} and \eqref{cprofile_sub} and the parametrized function  \eqref{envlope} with (red dashed line) and without (green dotted line) shift \eqref{shift}   in the supersonic (left panel) and subsonic regions (right panel).}}
	\label{fig_shift}
\end{figure}

 As a result of the asymptotic behaviors of \eqref{cprofile_sub} and \eqref{cprofile_sup}, the function $\eta_a$ and $\omega_a$ will be saturated to a constant value in the asymptotic regions where we summarize them as
\begin{subequations}
	\label{asy_sub}
	\begin{align}
		&	\begin{cases}
			&\eta_\text{sub}\rightarrow\eta_r=\sqrt{1-\omega_r^2/\omega^2}\\
			&\omega_\text{sub}\rightarrow \omega_r=\sqrt{2mc_r^2(1-M_r^2)\Omega}\\
			& M_\text{sub}\rightarrow M_r=v_r/c_r\\
		\end{cases} &\text{for}\quad z\rightarrow \infty,\\
		&	\begin{cases}
			&\eta_\text{sub}\rightarrow1\\
			&\omega_\text{sub}\rightarrow 0\\
						& M_\text{sub}\rightarrow 1\\
		\end{cases}& \text{for}\quad z\rightarrow -\infty,
	\end{align}
\end{subequations}
in the subsonic region and
\begin{subequations}
	\label{asy_sup}
	\begin{align}
		&	\begin{cases}
			&\eta_\text{sup}\rightarrow\eta_l=\sqrt{1+\omega_l^2/\omega^2}\\
			&\omega_\text{sup}\rightarrow \omega_l=\sqrt{2mc_l^2(M_l^2-1)\Omega}\\
						& M_\text{sup}\rightarrow M_l=v_l/c_l\\
		\end{cases} &\text{for}\quad z\rightarrow \infty,\\
		&	\begin{cases}
			&\eta_\text{sup}\rightarrow1\\
			&\omega_\text{sup}\rightarrow 0\\
						& M_\text{sup}\rightarrow 1\\
		\end{cases} &\text{for}\quad z\rightarrow -\infty.
	\end{align}
\end{subequations}
in the supersonic region.\\
{
To make \eqref{mode_eq2} analytically treatable, we approximate the first term of \eqref{mode_eq2} in a form
\begin{align}
	\left(M^2\eta^2\right)_\text{sub/sup}\simeq &\frac{1}{2} \Big[M_{r/l}^2\eta _{r/l}^2+1+\left(M_{r/l}^2\eta _{r/l}^2 -1\right)\no\\
	&\times \tanh (\kappa (z+\delta z_\text{sub/sup}))\Big]
	\label{envlope}
\end{align}
with the shift
\begin{align}
	&\delta z_\text{sub/sup}=\no\\
	&\frac{\left[\left(v_0^2-v_{r/l}^2\right) \left(c_{r/l}^2-v_0^2\right)\right] \left[2\omega^2+(3 v_0^2 +v_{r/l}^2)2m\Omega \right]}{4 v_0^2 \left(c_{r/l}^2-v_{r/l}^2\right) \left(\omega ^2+v_{r/l}^22m\Omega \right)},
	\label{shift}
	\end{align}
that the parametrized function matches not only at the asymptotical values but also  at $z=0$ (see Fig.~\ref{fig_shift}),
 although the shift does not affect the behaviors of the scattering processes asymptotically.
}

 Based upon the asymptotic regions of the  Penrose diagram (see Fig.~\ref{fig_penrose}), there are three incoming modes coming from the past horizon in the right (left) of the horizon $H_-^r$  ($H_-^l$) of the subsonic (supersonic) region and the past null infinity $I_-$ in the subsonic region with the following mode functions
\begin{align}
e^{-i \omega \tau}	 \varphi_{\omega,H}^{\text{in},r}(z)=\sqrt{\frac{1}{4\pi\omega}}e^{-i\omega\tau}e^{i \omega z}\qquad \text{from}\,\, H_-^r,
	\label{fhrin}
\end{align}
\begin{align}
e^{i \omega \tau}	\varphi_{\omega,H}^{\text{in},l}(z)=\sqrt{\frac{1}{4\pi\omega}}e^{i\omega\tau}e^{-i \omega z}\qquad \text{from}\,\, H_-^l,
	\label{fhlin}
\end{align}
\begin{align}
e^{-i \omega \tau}	\varphi_{\omega,I}^{\text{in},r}(z)=\sqrt{\frac{1}{4\pi\omega M_r\eta_r}}e^{-i\omega\tau}e^{-i \omega M_r\eta_r z}\qquad \text{from}\,\, I_-^r.
	\label{firin}
\end{align}
In the case of the metric given in (\ref{metric_conformal}), the mode functions shown above in the subsonic region ($v<c$) correspond to the positive frequency modes with respect to the time coordinate $\tau$ whereas the mode functions in the supersonic region ($v>c$) are the positive frequency modes with respect to the ``time'' coordinate $z$.
According to \eqref{mode_eq2}, the normalization of the mode functions $\varphi_{\omega}^{in}(z)$ as well as its frequency dependence of the mode functions can be chosen from the standard one $\varphi_{\omega}^{in}(z)= \frac{1}{\sqrt{4 \pi\omega}} e^{\pm i \omega z}$ by replacing $\omega \rightarrow {\omega M_\text{a}\eta_\text{a}}$ evaluated either at the horizon or the past null infinity, which is consistent with Refs.~ \cite{Fabbri2016,Roberto2019,Dudley2020}.
Similarly, there are three outgoing modes. They are  $u_{l}$ and $v_{l}$ modes  in the supersonic region toward the future null infinity in the respective asymptotic regions in terms of  the incoming and outgoing null coordinates $v_{l}=\tau+M_{l}z$ and $u_{l}=\tau-M_{l}z$ realized from the metric form \eqref{metric_conformal}.
Also, in the subsonic region, the  $u_{r}=\tau-M_{r}z$ mode is involved.  The mode functions are
\begin{align}
e^{-i\omega\tau}	\varphi_{\omega,u_r}^{\text{out},r}(z)=\sqrt{\frac{1}{4\pi\omega M_r \eta_r}}e^{-i\omega \tau}e^{i {\omega M_r\eta_r }z}\, \text{toward}\, I_+^r,
	\label{furout}
\end{align}
\begin{align}
e^{i\omega\tau}	\varphi_{\omega,u_l}^{\text{out},l}(z)=\sqrt{\frac{1}{4\pi\omega M_l \eta_l}}e^{i\omega \tau}e^{-i {\omega M_l\eta_l }z}\, \text{toward}\, I_{u,+}^{l},
	\label{fulout}
\end{align}
\begin{align}
e^{-i\omega\tau}	\varphi_{\omega,v_l}^{\text{out},l}(z)=\sqrt{\frac{1}{4\pi\omega M_l\eta_l}}e^{-i\omega\tau}e^{-i {\omega M_l\eta_l }z}\, \text{toward}\, I_{v,+}^{l}.
	\label{fvlout}
\end{align}
Notice that the correct choice of the normalization of the mode functions becomes essential to fulfill the unitarity conditions to be checked later.
The detailed mode functions will be determined in the next section.

\section{ solutions of the mode equations}\label{ABH3}

\subsection{Incoming mode $\theta_H^{\text{in},r}$}\label{subsec_fhr}
\begin{figure*}[t]
	\includegraphics[scale=0.55]{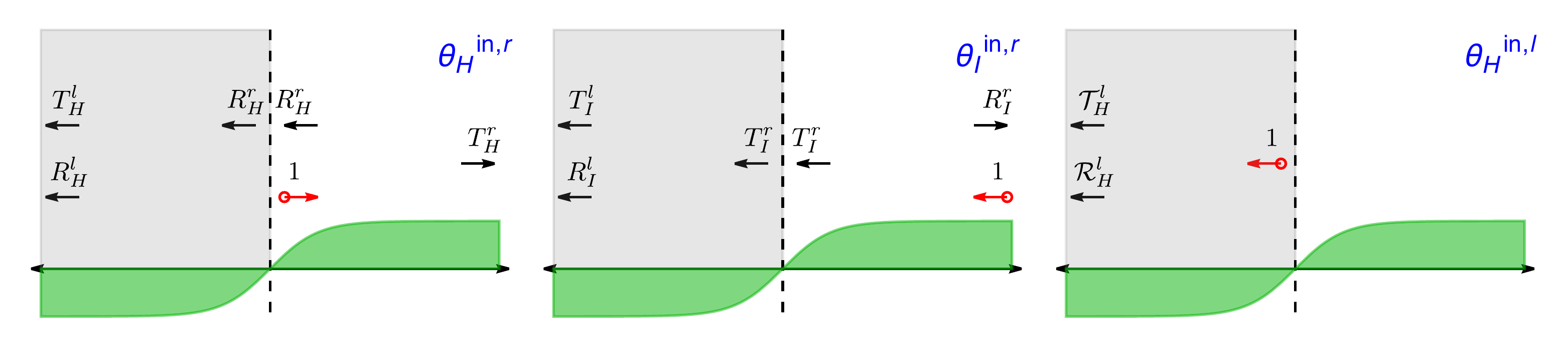}
	\caption{Scattering processes of the incoming mode $\theta_H^{\text{in},r}$(left),  $\theta_I^{\text{in},r}$(middle), and $\theta_H^{\text{in},l}$(right).  According to  \eqref{omega_a}, the green shadow is depicted for $+\omega^2_{\rm sub}(z)$ in the subsonic region and $-\omega^2_{\rm sup}(z)$ in the supersonic region. }
		\label{fig_fhr}
\end{figure*}
Firstly, we consider the incoming mode $\theta_H^{\text{in},r}$ from the right region of the past horizon  for $\omega>\omega_r$ by evaluating the threshold frequency $\omega_{\text sub}$ in the past horizon in the subsonic region shown in (\ref{asy_sub}) (see Fig.~\ref{fig_penrose}), which then partially reflects  back to the future horizon with the amplitude $R_H^r$ and partially  transmits  with the amplitude $T_H^r$ toward the future null infinity in the subsonic region ${I}_+^r$.
However, for  $\omega<\omega_r$, the incident mode will be totally reflected giving $T_H^r=0$ and $R_H^r=1$ toward the future horizon, which is referred as the boomerang trajectory of the gapped excitations \cite{Jannes2011}.
The scattering process can be schematically shown in Fig.~\ref{fig_fhr} (left) with the asymptotic behaviors of the mode function in a subsonic region ($x>0$) given by
\begin{widetext}
\begin{align}
	\varphi_{\omega,H}^{\text{in},r}(z)=
	\begin{cases}
		\sqrt{\frac{1}{4\pi \omega}} \exp{\left(i\omega z\right)}+R_H^r\sqrt{\frac{1}{4\pi \omega}} \exp{\left(-i{\omega}z\right)}, &  {z\rightarrow -\infty}.\\
		T_H^r\sqrt{\frac{1}{4\pi \omega M_r\eta_r}} \exp{\left(i{\omega M_r\eta_r}z\right)}, & {z\rightarrow \infty}\, .
	\end{cases}
	\label{phi_Hr}
\end{align}
\end{widetext}
The scattering coefficients in \eqref{phi_Hr} can be found from the approximate solution of \eqref{mode_eq2} with the parametrization function (\ref{envlope})
 together with the asymptotic values of $M_\text{sub}\eta_\text{sub}$ \eqref{asy_sub}. We choose one of two independent solutions  given by
\begin{align}
\varphi^\text{sub}_{\omega}(z)\propto& e^{-\frac{\pi  \omega }{2 \kappa }} \left(-e^{\kappa\, \delta z_\text{sub} }\right)^{\frac{i \omega }{\kappa }} \left(e^{\kappa  z}\right)^{\frac{i \omega }{\kappa }} \no\\
&\left(e^{2 \kappa  (z+\delta z_\text{sub})}+1\right)^{\frac{i \left(1-{ M_r\eta_r}\right) \omega }{2 \kappa}}\, _2F_1\left(a,b;c;y\right),
\label{phiHrsol}
\end{align}
where $_2F_1$ is the hypergeometric function with the arguments
\begin{subequations}
\begin{align}
	&a=\frac{i}{2} \frac{\left(1+M_r\eta_r\right)\omega}{\kappa},\\
	&b=1-\frac{i}{2} \frac{\left(1-M_r\eta_r\right)\omega}{\kappa},\\
	&c=1+\frac{i M_r\eta_r \omega  }{\kappa},\\
	& y=\frac{1}{1+e^{2 \kappa (z+\delta z_\text{sub}) }}\, ,
\end{align}
\end{subequations}
with $M_r\eta_r$in 	\eqref{asy_sub}.
 Since the hypergeometric function  $_2F_1(a,b,c;y=0)=1$ in the case of $1/\left(1+e^{2\kappa (z+\delta z_\text{sub})}\right) \rightarrow 0$ as $z \rightarrow \infty$, the amplitude in \eqref{phiHrsol} becomes an outgoing mode:
\begin{align}
	\varphi^\text{sub}_\omega(z) \propto e^{i \omega M_r\eta_r z}\qquad \text{as} \quad z\rightarrow \infty.
	\label{phi_asy}
\end{align}
Furthermore,  near the horizon ($z\rightarrow-\infty$), we use the identity
\begin{align}
_2F_1(a,b;c;y)=&\frac{\Gamma(c)\Gamma(c-a-b)}{\Gamma(c-a)\Gamma(c-b)}\no\\&\times{_2F_1}(a,b;a+b+1-c;1-y)\nonumber\\
&+\frac{\Gamma(c)\Gamma(a+b-c)}{\Gamma(a)\Gamma(b)}(1-y)^{c-a-b}\no\\&\times{_2F_1}(c-a,c-b;1+c-a-b;1-y)
\label{transformation}
\end{align}
to rewrite \eqref{phiHrsol} as
\begin{align}
	\varphi^\text{sub}_\omega(z)\propto&\frac{\Gamma(c)\Gamma(c-a-b)}{\Gamma(c-a)\Gamma(c-b)}e^{i\omega z}+\frac{\Gamma(c)\Gamma(a+b-c)}{\Gamma(a)\Gamma(b)}e^{-i\omega z}\no\\
	&\text{as} \quad z\rightarrow -\infty
	\label{phi_hor}
\end{align}
with $_2F_1(a,b,c;1-y=0)=1$ again since $1-y=e^{2\kappa z}/(1+e^{2\kappa z})\rightarrow 0$ for $z\rightarrow -\infty$.
 Comparing \eqref{phi_Hr}, \eqref{phi_asy} and \eqref{phi_hor}, we are able to extract the reflection coefficient $R_H^r$ and the transmission coefficient $T_H^r$ obtained as
\begin{align}
	&R_H^r=\frac{\Gamma(c-a)\Gamma(c-b)\Gamma(a+b-c)}{\Gamma(a)\Gamma(b)\Gamma(c-a-b)},\\
	&T_H^r=\sqrt{M_r\eta_r}\frac{\Gamma(c-a)\Gamma(c-b)}{\Gamma(c)\Gamma(c-a-b)}.
\end{align}
Consequently they are
\begin{align}
		T_H^r&=\sqrt{M_r\eta _r }\frac{ \Gamma \left(\scriptstyle\frac{i   \left(1+{M_r \eta _r}\right)\omega}{2 k}\right) \Gamma \left(\scriptstyle1+\frac{i   \left(1+{M_r \eta _r}\right)\omega}{2 k}\right)}{\Gamma \left(\scriptstyle\frac{i \omega }{k}\right) \Gamma \left(\scriptstyle1+\frac{i   M_r \eta _r\omega}{k}\right)}\theta(\omega-\omega_r)\label{thr}\,,
	\end{align}
	\begin{subequations}\label{rhr}
		\begin{align}
			R_H^r=&\frac{\Gamma \left(\scriptstyle-\frac{i \omega }{\kappa }\right) \Gamma \left(\scriptstyle\frac{i   \left(1+M_r \eta _r\right)\omega}{2 \kappa }\right) \Gamma \left(\scriptstyle1+\frac{i \left(1+M_r \eta _r \right)\omega}{2 \kappa }\right)}{\Gamma \left(\scriptstyle\frac{i \omega }{\kappa }\right) \Gamma \left(\scriptstyle-\frac{i  \left(1-M_r \eta _r\right)\omega}{2 \kappa }\right) \Gamma \left(\scriptstyle1-\frac{i   \left(1-M_r \eta _r\right)\omega}{2 \kappa }\right)},
			\quad \omega>\omega_r\,,\\
			&=1,\quad \omega\le\omega_r \, ,
		\end{align}
\end{subequations}
which satisfy the unitarity relation
\begin{align}
	\vert R_H^r\vert^2+	\vert T_H^r\vert^2=1.
\end{align}

In \eqref{thr}, $\theta(\omega-\omega_r)$ is the Heaviside step function.
Afterward, the  mode function  with the amplitude $R_H^r$  that propagates from the subsonic region to the supersonic region  transmits (reflects) to the future null infinity in the supersonic region as in terms of the $v$ ($u$) mode with the following asymptotic behaviors:
\begin{widetext}
\begin{align}
	\varphi_{\omega,H}^{\text{in},r}=
	\begin{cases}
		R_H^r\sqrt{\frac{1}{4\pi \omega}} \exp{\left(i\omega z\right)}, & {z\rightarrow -\infty},\\
		 T_H^l\sqrt{\frac{1}{4\pi \omega M_l\eta_l}}\exp{\left(-i{\omega M_l\eta_l}z\right)}+R_H^l\sqrt{\frac{1}{4\pi \omega M_l\eta_l}} \exp{\left(i{\omega M_l\eta_l}z\right)}, &  {z\rightarrow \infty}.
	\end{cases}
	\label{phi_Hl}
\end{align}
\end{widetext}
 For the supersonic region, the  solution
has the general solution
\begin{align}
	\varphi_\omega^\text{sup}(z)\propto	&e^{-\frac{\pi  \omega }{2 \kappa }} \left(-e^{ \kappa \delta z}\right)^{\frac{i \omega }{\kappa }} \left(e^{\kappa  z}\right)^{\frac{i \omega }{\kappa }} \no\\&\left(e^{(\kappa  (z+\delta z_\text{sup}))}+1\right)^{-\frac{i  \left(M_l\eta _l -1\right)\omega}{ \kappa }}\, _2F_1\left(a,b;c;y\right)
		\label{phiHlsol}
\end{align}
with the arguments
\begin{subequations}
		\begin{align}
			&a=-\frac{i}{2} \frac{\left(M_l\eta_l-1\right)\omega}{\kappa},\\
			&b=1-\frac{i}{2} \frac{\left(M_l\eta_l-1\right)\omega}{\kappa},\\
			&c=1+\frac{i \omega  }{\kappa},\\
			& y=\frac{1}{1+e^{-2 \kappa (z+\delta z) }}\, ,
		\end{align}
\end{subequations}
where $M_l\eta_l$ is in \eqref{asy_sup}.
For $z\rightarrow -\infty$, giving $y\rightarrow 0$, where $_2F_1(a,b;c;0)=1$, the solution \eqref{phiHlsol} correctly describes the incident wave  from the future horizon as
\begin{align}
		\varphi_\omega^\text{sup}(z)\propto  e^{ -i\omega z}\qquad \text{as} \quad z\rightarrow -\infty.
		\label{phiHlsol2}
\end{align}
However, for $z\rightarrow \infty$, we use the transformation
\eqref{transformation} to rewrite \eqref{phiHlsol} as
\begin{align}
	\varphi_\omega^\text{sup}(z)\propto&  \frac{\Gamma(c)\Gamma(c-a-b)}{\Gamma(c-a)\Gamma(c-b)}e^{-i\omega\eta_lM_l z}+\frac{\Gamma(c)\Gamma(a+b-c)}{\Gamma(a)\Gamma(b)}\no\\
	&e^{i\omega\eta_l M_lz}\qquad \text{as} \quad z\rightarrow \infty.
	\label{phiHlsol3}
\end{align}
Comparing \eqref{phi_Hl} and \eqref{phiHlsol3} enables us to extract the scattering coefficients
\begin{align}
		T_H^l=&\sqrt{ M_l \eta _l}\frac{\Gamma \left(\scriptstyle1+\frac{i \omega }{\kappa }\right)  \Gamma \left(\scriptstyle\frac{i  M_l \eta _l\omega}{\kappa }\right)}{\Gamma \left(\scriptstyle\frac{i   \left(M_l \eta _l+1\right)\omega}{2 \kappa }\right) \Gamma \left(\scriptstyle1+\frac{i \left(M_l \eta _l  + \right)\omega}{2 \kappa }\right)}R_H^r,\label{Thl}
	\end{align}
	\begin{align}
		R_H^l=\sqrt{ M_l \eta _l}\frac{\Gamma \left(\scriptstyle1+\frac{i \omega }{\kappa }\right)  \Gamma \left(\scriptstyle-\frac{i \omega  M_l \eta _l}{\kappa }\right)}{\Gamma \left(\scriptstyle-\frac{i \omega  \left(M_l \eta _l-1\right)}{2 \kappa }\right) \Gamma \left(\scriptstyle1-\frac{i \omega  \left(M_l \eta _l-1\right)}{2 \kappa }\right)}R_H^r.\label{Rhl}
\end{align}
For the whole scattering process from the mode function $\varphi^{\text{in},r}_{\omega,H}(z)$, the above coefficients satisfy the relation
\begin{align}
	\vert T_H^l\vert^2-\vert R_H^l\vert^2+\vert T_H^r\vert^2=1,
	\label{conservation_hr}
\end{align}
where the minus sign is due to the negative norm of the reflected $u$ mode in the supersonic region.\\

\subsection{Incoming mode $\theta_I^{\text{in},r}$}\label{subsec_fir}
The second incoming propagating mode  $\theta_I^{\text{in},r}$ is also considered  for $\omega>\omega_r$  from past null infinity ${I}_-^r$ in the subsonic region, which scatters into the reflected mode with the reflection coefficient ${R}_I^r$ in the future null infinity, and the transmitted mode with the transmission coefficient ${T}_I^r$ toward the future horizon [see Figs.~\ref{fig_penrose} and \ref{fig_fhr} (middle)].
\begin{widetext}
The incoming mode $\theta_I^{\text{in},r}(\tau,z)$ has the asymptotic forms in the subsonic region as
\begin{align}
	\varphi_{\omega,I}^{\text{in},r}=
	\begin{cases}
		T_I^r\sqrt{\frac{1}{4\pi \omega}} \exp{\left(-i\omega z\right)}, & {z\rightarrow -\infty},\\
		\sqrt{\frac{1}{4\pi \omega M_r\eta_r}}\exp{\left(-i{\omega M_r\eta_r}z\right)}+R_I^r\sqrt{\frac{1}{4\pi \omega M_r\eta_r}} \exp{\left(i{\omega M_r\eta_r}z\right)}, &  {z\rightarrow \infty}.
	\end{cases}
	\label{phi_Ir}
\end{align}
Then,
in the supersonic region, the mode with the amplitude ${T}_I^r$ coming from the future horizon scatters into the  transmitted (reflected) mode to the future null infinity in the supersonic region in terms of the $v$ ($u$) mode with the following asymptotic behaviors
\begin{align}
	\varphi_{\omega,I}^{\text{in},r}=
	\begin{cases}
		T_I^r\sqrt{\frac{1}{4\pi \omega}} \exp{\left(-i\omega z\right)}, & {z\rightarrow -\infty},\\T_I^l
		\sqrt{\frac{1}{4\pi \omega M_l \eta_l}}\exp{\left(-i{\omega M_l\eta_l}z\right)}+R_I^l\sqrt{\frac{1}{4\pi \omega M_l\eta_l}} \exp{\left(i{\omega M_l\eta_l}z\right)}, &  {z\rightarrow \infty}.
	\end{cases}
	\label{phi_Il}
\end{align}
\end{widetext}
Following the same procedure as in Sec.~\ref{subsec_fhr}  to match the incoming mode $\varphi_{\omega,I}^{\text{in},r}$ of the  solutions with the mode function $\varphi_\omega^\text{sub/sup}$ near the horizon and the asymptotic regions, one obtains the  scattering coefficients
\begin{align}
		&T_I^r=\frac{\Gamma \left(\scriptstyle\frac{i   \left(M_r \eta _r+1\right)\omega}{2 \kappa }\right) \Gamma \left(\scriptstyle1+\frac{i \left(M_r \eta _r  +1 \right)\omega}{2 \kappa }\right)}{\sqrt{M_r \eta _r} \Gamma \left(\scriptstyle1+\frac{i \omega }{\kappa }\right) \Gamma \left(\scriptstyle\frac{i \omega  M_r \eta _r}{\kappa }\right)}\theta(\omega-\omega_r),\label{TIr}\\
		&R_I^r=\frac{\Gamma \left(\scriptstyle-\frac{i \omega  M_r \eta _r}{\kappa }\right) \Gamma \left(\scriptstyle\frac{i   \left(1+M_r \eta _r\right)\omega}{2 \kappa }\right) \Gamma \left(\scriptstyle1+\frac{i \left(1+M_r \eta _r  \right)\omega}{2 \kappa }\right)}{\Gamma \left(\scriptstyle\frac{i \omega  M_r \eta _r}{\kappa }\right) \Gamma \left(\scriptstyle-\frac{i   \left(M_r \eta _r-1\right)\omega}{2 \kappa }\right) \Gamma \left(\scriptstyle1-\frac{i   \left(M_r \eta _r-1\right)\omega}{2 \kappa }\right)}\no\\
		&\quad\qquad\times\theta(\omega-\omega_r),\label{RIr}\\
		&R_I^l= \sqrt{ M_l\eta _l} \frac{\Gamma \left(\scriptstyle1+\frac{i \omega }{\kappa }\right)\Gamma \left(\scriptstyle-\frac{i \omega  M_l \eta _l}{\kappa }\right)}{\Gamma \left(\scriptstyle-\frac{i   \left(M_l \eta _l-1\right)\omega}{2 \kappa }\right) \Gamma \left(\scriptstyle1-\frac{i   \left(M_l \eta _l-1\right)\omega}{2 \kappa }\right)}T_I^r,\label{RIl}\\
		&T_I^l= \sqrt{M_l\eta _l }\frac{\Gamma \left(\scriptstyle1+\frac{i \omega }{\kappa }\right) \Gamma \left(\scriptstyle\frac{i \omega  M_l \eta _l}{\kappa }\right)}{\Gamma \left(\scriptstyle\frac{i   \left(M_l \eta _l+1\right)\omega}{2 \kappa }\right) \Gamma \left(\scriptstyle1+\frac{i \left(M_l \eta _l  +1 \right)\omega}{2 \kappa }\right)}T_I^r.\label{TIl}
\end{align}
These coefficients obey the relations
\begin{align}
	&\vert R_I^r\vert^2+\vert T_I^r\vert^2=1,\\
	&\vert T_I^l\vert^2-\vert R_I^l\vert^2+\vert R_I^r\vert^2=1.
\end{align}
Notice that this scattering process can only happen as $\omega>\omega_r$. Again, the negative norm of the reflected $u$ mode in the supersonic region gives the minus sign in  the equation.

\subsection{Incoming mode $\theta_H^{\text{in},l}$}
The third incoming mode under consideration is $\theta_H^{\text{in},l}$, which is incident from the  past horizon in the supersonic region. As seen in Figs.~\ref{fig_penrose} and \ref{fig_fhr} (right), we have one partial reflected mode, in particular  of the  negative norm state accompanying with reflection coefficient $\mathcal{R}_H^l$, and one partial transmitted mode with the transmission coefficient $\mathcal{T}_H^l$. Note that this scattering process has nothing to do with the subsonic region due to the formation of the analog horizon where the modes are forced to  travel toward $x\rightarrow -\infty$ away from the  horizon and will not escape into the subsonic region.

The asymptotic behaviors of the incoming mode $\varphi_{\omega,H}^{\text{in},l}$ in the supersonic region are
\begin{widetext}
\begin{align}
	\varphi_{\omega,H}^{\text{in},l}=
	\begin{cases}
		\sqrt{\frac{1}{4\pi \omega}} \exp{\left(-i\omega z\right)}, & {z\rightarrow -\infty},\\
		\mathcal{T}_H^l\sqrt{\frac{1}{4\pi \omega M_l \eta_l}}\exp{\left(-i{\omega M_l\eta_l}z\right)}+\mathcal{R}_H^l\sqrt{\frac{1}{4\pi \omega M_l \eta_l}} \exp{\left(i{\omega M_l\eta_l}z\right)}, &  {z\rightarrow \infty}.
	\end{cases}
	\label{phi_H_l}
\end{align}
\end{widetext}
It is straightforward to obtain the  scattering coefficients by letting $T_I^r=1$ in \eqref{RIl} and \eqref{TIl} as
	\begin{align}
		&\mathcal{R}_H^l=\sqrt{ M_l \eta _l}\frac{\Gamma \left(\scriptstyle1+\frac{i \omega }{\kappa }\right)  \Gamma \left(\scriptstyle-\frac{i \omega  M_l \eta _l}{\kappa }\right)}{\Gamma \left(\scriptstyle-\frac{i \omega  \left(M_l \eta _l-1\right)}{2 \kappa }\right) \Gamma \left(\scriptstyle1-\frac{i \omega  \left(M_l \eta _l-1\right)}{2 \kappa }\right)},\label{mRhl}\\
		&\mathcal{T}_H^l=\sqrt{ M_l \eta _l}\frac{\Gamma \left(\scriptstyle1+\frac{i \omega }{\kappa }\right)  \Gamma \left(\scriptstyle\frac{i \omega  M_l \eta _l}{\kappa }\right)}{\Gamma \left(\scriptstyle\frac{i   \left(M_l \eta _l+1\right)\omega}{2 \kappa }\right) \Gamma \left(\scriptstyle1+\frac{i \left(M_l \eta _l  + 1\right)\omega}{2 \kappa }\right)},\label{mThl}
\end{align}
which satisfy the relation
\begin{align}
 \vert \mathcal{T}_H^l \vert^2-\vert \mathcal{R}_H^l \vert^2=1.
 \label{conservation_hl}
\end{align}
In the paper of Ref.~\cite{Fabbri2016} where the gapless cases are under consideration, the scattering coefficients  are shown only in the subsonic region where our results are consistent with them in the limit of $\eta_{r} \rightarrow 1 ~({\Omega \rightarrow 0})$.

\section{Construction of  $S$-matrix and Bogoliubov transformation}\label{ABH4}
Having all the scattering coefficients from all incoming modes, we are able to construct the mode expansion of the field operator in terms of incoming modes
\begin{align}
	\delta\hat{\theta}(\tau,z)=&\int d\omega  \Bigg\{ e^{-i\omega\tau} \Big[ \hat{a}_{\omega,H}^{\text{in},r} \,\varphi_{\omega,H}^{\text{in},r}+\hat{a}_{\omega,I}^{\text{in},r} \,\varphi_{\omega,I}^{\text{in},r}\no\\
	&+\left(\hat{a}_{\omega,H}^{\text{in},l}\right)^\dagger \,\varphi_{\omega,H}^{\text{in},l\,\ast}\Big]+\text{H.c.}\Bigg\},
	\label{mode_exp}
\end{align}
or in terms of outgoing modes
\begin{align}
	\delta\hat{\theta}(\tau,z)=&\int d\omega\Bigg\{ e^{-i\omega\tau} \Big[ \hat{a}_{\omega,u_r}^{\text{out},r} \,\varphi_{\omega,u_r}^{\text{out},r}+\hat{a}_{\omega,v_l}^{\text{out},l} \,\varphi_{\omega,v_l}^{\text{out},l}\no\\
	&+\left(\hat{a}_{\omega,u_l}^{\text{out},l}\right)^\dagger \,\varphi_{\omega,u_l}^{\text{out},l\,\ast}\Big]+\text{H.c.}\Bigg\},
	\label{mode_exp2}
\end{align}
according to the asymptotic states defined on the boundaries of the Penrose diagram in Fig.~\ref{fig_penrose}. The relation between incoming and outgoing modes given by (\ref{fhrin})-(\ref{fvlout}) can be summarized into the $S$ matrix  to be
\begin{align}
	\begin{pmatrix}
		\varphi_{\omega,H}^{\text{in},r}\\
		\varphi_{\omega,H}^{\text{in},l\ast}\\
		\varphi_{\omega,I}^{\text{in},r}
	\end{pmatrix}=S\cdot \begin{pmatrix}
	\varphi_{\omega,u_r}^{\text{out},r}\\
	\varphi_{\omega,u_l}^{\text{out},l\ast}\\
	\varphi_{\omega,v_l}^{\text{out},l}
\end{pmatrix},
\end{align}
where
\begin{align}
	S=
	\begin{pmatrix} S_{ur,Hr} & S_{ul,Hr} & S_{vl,Hr}\\
		S_{ur,Hl} & S_{ul,Hl} & S_{vl,Hl}\\
		S_{ur,Ir} & S_{ul,Ir} & S_{vl,Ir}
	\end{pmatrix}\,.
\label{bogo_trans}
\end{align}
The subscript of the element $S_{i,j}$ indicates the relation between the $i$ outgoing mode and the $j$ incoming mode. Substituting \eqref{bogo_trans} into \eqref{mode_exp} and comparing with \eqref{mode_exp2},  the Bogoliubov transformation can be read off as
\begin{align}\label{Bogoliubov_1}
		\begin{pmatrix}
		\hat{a}_{\omega,u_r}^{\text{out},r}\\
		(\hat{a}_{\omega,u_l}^{\text{out},l})^\dagger\\
		\hat{a}_{\omega,v_l}^{\text{out},l}
	\end{pmatrix}=
\begin{pmatrix} S_{ur,Hr} & S_{ur,Hl} & S_{ur,Ir}\\
	S_{ul,Hr} & S_{ul,Hl} & S_{ul,Ir}\\
	S_{vl,Hr} & S_{vl,Hl} & S_{vl,Ir}
\end{pmatrix}
\cdot \begin{pmatrix}
		\hat{a}_{\omega,H}^{\text{in},r}\\
		(\hat{a}_{\omega,H}^{\text{in},l})^\dagger\\
		\hat{a}_{\omega,I}^{\text{in},r}
	\end{pmatrix},
\end{align}
where  the $S$-matrix elements are related to all above transmission and reflection coefficients below:
\begin{subequations}
\begin{align}
	&S_{ur,Hr}=T_H^r,\quad S_{ur,Hl}=0,\quad S_{ur, Ir}=R_I^r,\\
	&S_{ul,Hr}=R_H^l,\quad S_{ul,Hl}=\mathcal{T}_H^{l\ast},\quad S_{ul, Ir}=R_I^l,\\
	&S_{vl,Hr}=T_H^l,\quad S_{vl,Hl}=\mathcal{R}_H^{l\ast},\quad S_{vl, Ir}=T_I^l.
\end{align}
\end{subequations}

When studying a physical effect of some quantum field in a curved spacetime, an important
 step is the identification of a quantum state or a class of quantum states which adequately
describes the given physical situation.
Based upon the mode expansion in (\ref{mode_exp2}) the natural vacuum state can be defined as  the  Boulware state, which is annihilated by the annihilation operators $ \hat{a}_{\omega,I}^{\text{in},r}, \hat{a}_{\omega,H}^{\text{in},r}$ and $\hat{a}_{\omega,H}^{\text{in},l}$, namely
\begin{align}
 \hat{a}_{\omega,I}^{\text{in},r}\vert B\rangle=0\quad \hat{a}_{\omega,H}^{\text{in},r}\vert B\rangle=0, \quad \text{and} \quad \hat{a}_{\omega,H}^{\text{in},l}\vert B\rangle=0 \,.
\end{align}
Here we consider the Unruh state, a stationary state that can
be thought of as describing a hot body, namely the black hole, immersed in vacuum. In particular,
it contains no particles coming from the past null infinity, while at the future null infinity,  the particles can be produced. This is consistent with blackbody radiation at the Hawking temperature. Thus the Unruh state
is generally considered to be the appropriate state for the description of
gravitational collapse of the black hole \cite{Unruh1981}.
The positive and negative frequency modes of the Unruh state are defined with respect to the Kruskal time
$U=\pm e^{-\kappa (\tau-z)}/\kappa$ where $+(-)$ corresponds to supersonic (subsonic) region of the horizon with the mode function
\begin{equation}
		\varphi_H^K=\sqrt{\frac{1}{4\pi\omega_K}}e^{-i\omega_KU}\, .
\end{equation}
Thus,  the field operator can also be expanded in terms of the mode functions $\varphi_H^K$ and $\varphi_I^{\text{in},r}$:
\begin{align}
\delta	\hat{\theta}=&\int d\omega_K\left(\hat{a}_{\omega_K}\varphi_H^K+\hat{a}_{\omega_K}^\dagger\varphi_H^{K\ast}\right)\no\\
&+\int d\omega\left(\hat{a}_{\omega,I}^{\text{in},r}\,\varphi_{\omega,I}^{\text{in},r}+(\hat{a}_{\omega,I}^{\text{in},r})^\dagger\,\varphi_{\omega,I}^{\text{in},r\ast}\right),
\end{align}
where the  Unruh state is annihilated by the annihilation operators $\hat{a}_{\omega_K}, \hat{a}_{\omega_K}^\dagger$:
\begin{align}
	\hat{a}_{\omega_K}\vert U\rangle=0\qquad \text{and} \quad \hat{a}_{\omega,I}^{\text{in},r}\vert U\rangle=0 \,.
\end{align}

The Bogoliubov transformations between two sets of the creation and annihilation operators are expressed as
\begin{align}\label{Bogoliubov_2}
	&\hat{a}_H^{\text{in},r}=\int d\omega_K\left[\alpha_{\omega_K,\omega}^r\hat{a}_{\omega_K}+\beta_{\omega_K,\omega}^{r\ast}\hat{a}_{\omega_K}^\dagger\right],\nonumber\\
	&\hat{a}_H^{\text{in},l}=\int d\omega_K\left[\alpha_{\omega_K,\omega}^l\hat{a}_{\omega_K}+\beta_{\omega_K,\omega}^{l\ast}\hat{a}_{\omega_K}^\dagger\right]
\end{align}
with the coefficients \cite{Anderson2013}
\begin{subequations}	\label{bogo_coe}
\begin{align}
	 \alpha_{\omega_K,\omega}^r&=\frac{1}{2\pi\kappa}\sqrt{\frac{\omega}{\omega_K}}(-i\omega_K)^{i\omega/\kappa}\Gamma\left(\frac{-i\omega}{\kappa}\right),\\
	 \beta_{\omega_K,\omega}^r&=\frac{1}{2\pi\kappa}\sqrt{\frac{\omega}{\omega_K}}(-i\omega_K)^{-i\omega/\kappa}\Gamma\left(\frac{i\omega}{\kappa}\right),\\
	 \alpha_{\omega_K,\omega}^l&=\frac{1}{2\pi\kappa}\sqrt{\frac{\omega}{\omega_K}}(i\omega_K)^{-i\omega/\kappa}\Gamma\left(\frac{i\omega}{\kappa}\right),\\
	 \beta_{\omega_K,\omega}^l&=\frac{1}{2\pi\kappa}\sqrt{\frac{\omega}{\omega_K}}(i\omega_K)^{i\omega/\kappa}\Gamma\left(\frac{-i\omega}{\kappa}\right).
\end{align}
\end{subequations}
\begin{figure}[t]
	\includegraphics[width=0.8\columnwidth]{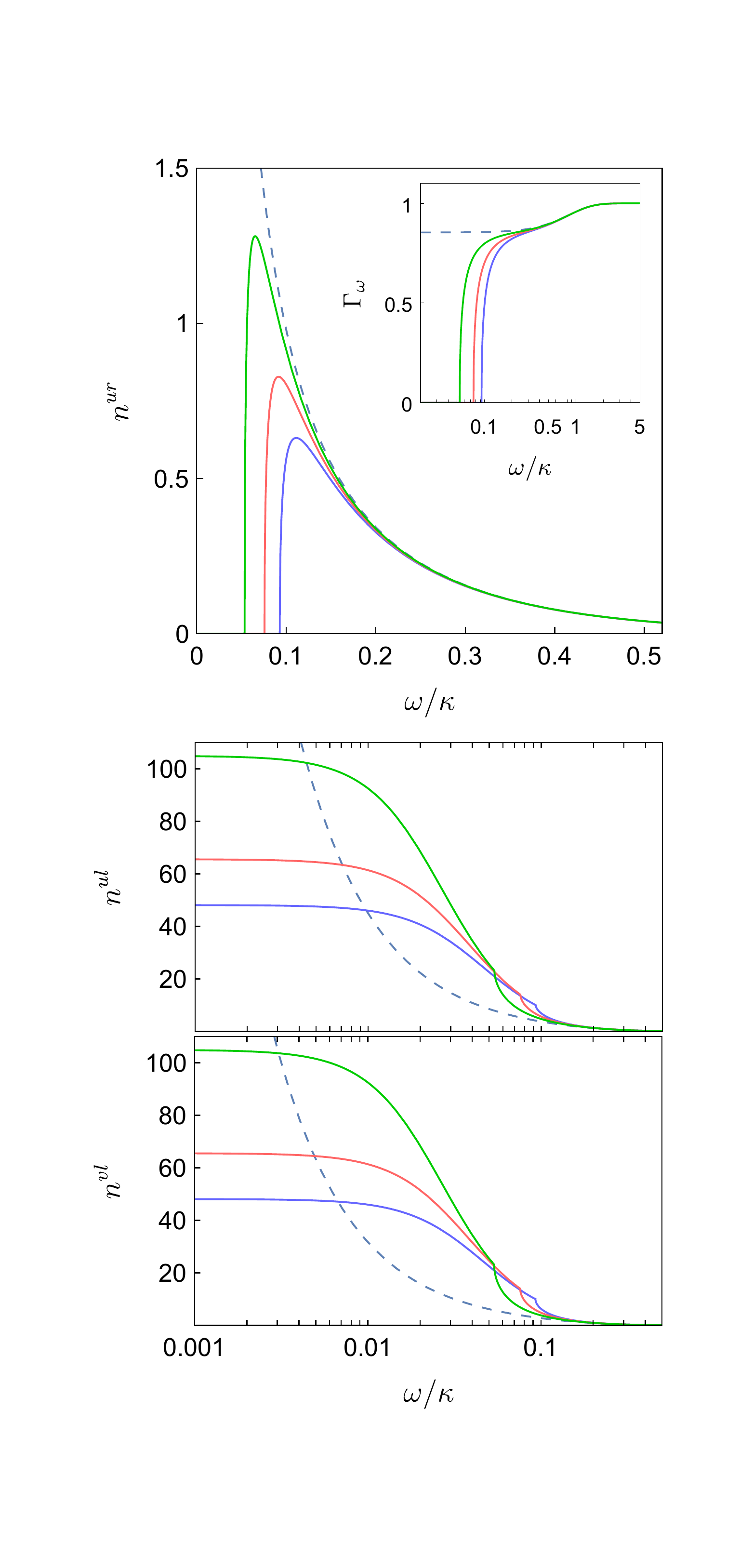}
	\caption{Plot of the spectrum $n^{ur},\,n^{ul},$ and $n^{vl}$ in (\ref{nur}), (\ref{nul}) and (\ref{nvl}) as the function of $\omega/\kappa$ with various values of the Rabi-frequency $\Omega/\kappa=\,0$ (dashed line), $0.001$ (green line), $0.002$ (red line), and $0.003$ (blue line). Notice that  the comparison is made between the gapped cases ($\Omega\neq0$) and the gapless cases ($\Omega=0$).  The inset in $n^{ur}$ shows the graybody factor $\Gamma_\omega$ in (\ref{graybody}). The parameters are the same as those  in Fig. \ref{fig_c}.}
	\label{fig_nur}
\end{figure}
 After introducing the appropriate quantum state, we are ready to compute the particle densities of each mode produced from the Unruh state as well as the mode mixing due to the existence of the negative norm state and their mode correlators.

\section{particle densities and mode correlators}\label{ABH5}
We first calculate the particle density of the $\varphi_{\omega,u_r}^{\text{out},r}$ mode in the subsonic region, an analog of the Hawking mode using the Bogoliubov transformations (\ref{Bogoliubov_1}) and (\ref{Bogoliubov_2}), which can be expressed as the thermal spectrum
\begin{align}
	n^{ur}&=\langle U\vert (\hat{a}^{\text{out},r}_{\omega,u_r})^\dagger\,\hat{a}^{\text{out},r}_{\omega,u_r}\vert U\rangle\nonumber\\
	&=\vert T_H^r\vert^2\frac{1}{e^{\frac{\omega}{T_\text{hw}}}-1}=\frac{\Gamma_\omega}{e^{\frac{\omega}{T_\text{hw}}}-1}
	\label{nur}
\end{align}
with the Hawking temperature $T_\text{hw}=\kappa/2\pi$ even for the gapped excitations shown in Fig. \ref{fig_nur}.
The accompanying  graybody factor is obtained as
\begin{align}
	\Gamma_\omega=&\frac{\sinh \left(\frac{\pi  \omega }{\kappa}\right) \sinh \left(\frac{\pi M_r \eta_r \omega }{\kappa }\right) }{\sinh^2\left(\frac{\pi  \omega  \left(1+M_r\eta_r  \right)}{2 \kappa}\right)}.
	\label{graybody}
\end{align}
\begin{figure*}[th]
	\includegraphics[scale=0.5]{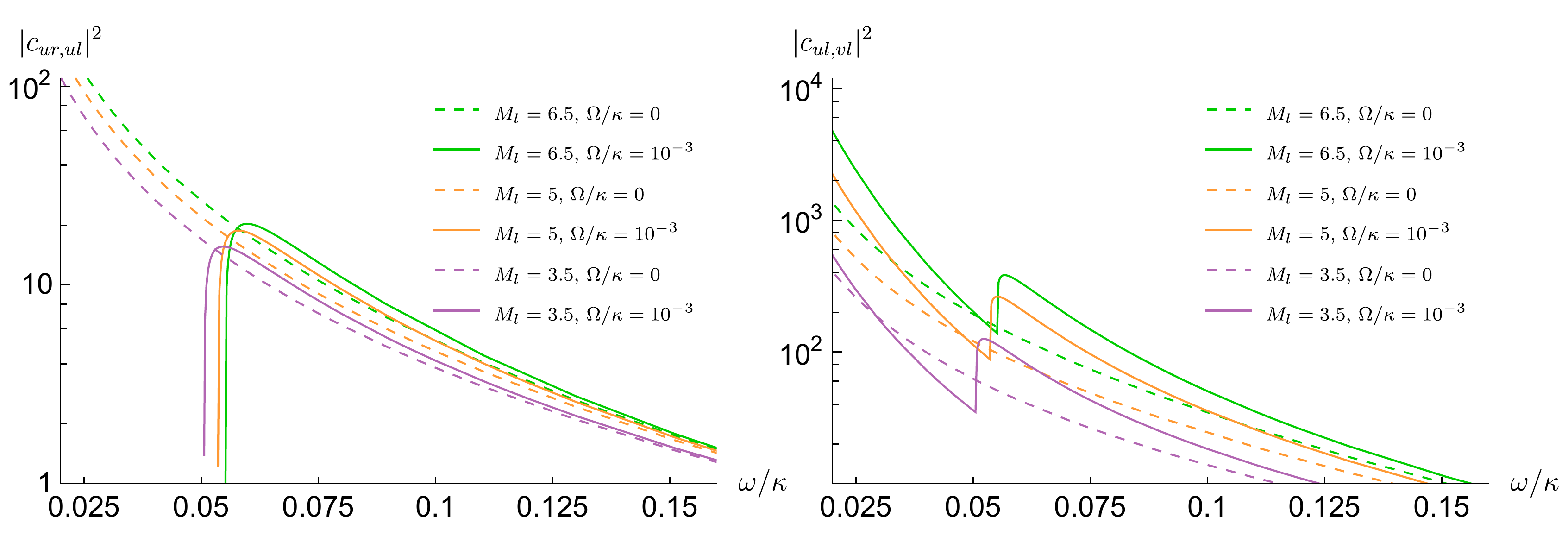}
	\caption{The magnitude of $|c_{ur,ul}|^2$ and  $|c_{ul,vl}|^2$ in (\ref{curul}) and (\ref{cvlul}) are shown in the left and right panels, respectively. We compare the gapped cases (solid lines) with gapless cases (dashed lines) for various values of Mach number $M_l=6.5$ (green line), $5$ (orange line), and $3.5$ (purple line). The other parameters are fixed as $c_l=0.4,c_r=1.34$. }
	\label{fig_ulur}
\end{figure*}
The produced particles are mainly due to the introducing Unruh state in the past horizon in the $r$ region where the modes can partially transmit to the future null infinity, giving the analog Hawking radiation.
 The obtained graybody factor above returns to the expression of the gapless case in Ref.~\cite{Fabbri2016} in the limits of  $\eta_r=1$ ($\Omega=0$).
For small frequency $\omega\ll \kappa$, the graybody factor can be approximated as
$4 M_r \eta _r/\left(1+M_r \eta _r\right)^2$
which is the same as the graybody factor  in the case of  the steplike change of the sound speed in Ref.~\cite{syu2022} with the constant velocity set to be $v=v_r$ where the dispersive effects can be ignored in such small frequencies.   The main influences of the gap energy  $m_\text{eff}$ in \eqref{KGE} to the graybody factor \eqref{graybody} can be seen from the existence of  a threshold $\omega_r$ where $n^{ur}=0$ if $\omega<\omega_r$  shown in the inset of Fig.~\ref{fig_nur}.  This is just the critical value of frequency, below which in the subsonic region
 the modes from the past horizon in the $r$ region will be totally reflected  toward the future horizon, giving no Hawking radiation in the future null infinity shown in Fig.~\ref{fig_fhr}. This can be realized to rewrite (\ref{mode_eq2}) as the time-independent Schr\"odinger-like equation where $\omega^2_\text{a}$ plays a role as the  effective potential term. The relative  large value of $m_{\rm eff}$ in the subsonic region driven by the large Rabi coupling constant $\Omega$ gives the relatively large value of $\omega^2_{\rm sub}$ (\ref{omega_a}) that leads to the large scattering effects, giving smaller transmission coefficient $T_H^r$ and thus resulting in the smaller graybody factor. When $\omega \rightarrow \infty$, $\Gamma_\omega \rightarrow 1$ as expected.

Next, we consider particle spectrum of the modes inside the analog horizon, namely $\varphi_{\omega,v_l}^{\text{out},l}$ and $\varphi_{\omega,u_l}^{\text{out},l}$. Let us first study the
behavior of the $\varphi_{\omega,u_l}^{\text{out},l}$ mode of the negative norm state, which is also called the partner of the Hawking mode.
The particle spectrum is obtained as
\begin{align}
	n^{ul}&=\langle U\vert (\hat{a}^{\text{out},l}_{\omega,u_l})^\dagger\,\hat{a}^{\text{out},l}_{\omega,u_l}\vert U\rangle\nonumber\\
	&=\frac{1}{e^{\frac{\omega}{T_\text{hw}}}-1}\big\vert  e^{\frac{\omega}{2T_\text{hw}}}R_H^l+ \mathcal{T}_H^{l\,\ast}\big\vert^2+\big\vert R_I^l\big\vert^2
		\label{nul}
\end{align}
where the coefficients can be substituted from \eqref{Rhl}, \eqref{RIl} and \eqref{mThl}. Although there is no particle coming from the past null infinity, the nature of the negative norm state of $\varphi_{\omega,u_l}^{\text{out},l}$ in the supersonic region gives rise to vacuum instabilities due to the mode mixing, triggering the particle production due to the contribution of $ R_I^l $. In addition, the scattering of the modes from the  Unruh state in the past horizon (in the $l$ and $r$ regions) contributes  particle production due to the coefficients of ${R}_H^l$ and $\mathcal{T}_H^l$. The net result of the particle spectrum exhibits nonthermal.
Figure~\ref{fig_nur} shows the particle density $n^{ul}$ does not change smoothly across $\omega_r$. The main reason is due to the fact that
 the modes coming from the past horizon in the $r$ region will totally be reflected to the future horizon  when $\omega <\omega_r$, giving the enhancement of the particle production of $n^{ul}$ as compared with the modes with frequencies $\omega > \omega_r$. Also, below threshold frequency $\omega_r$, there does not exist the propagating incoming mode originally from the past null infinity in the subsonic region shown in Fig.~\ref{fig_fhr} (middle), giving the vanishing of   $ R_I^l $ when $\omega <\omega_r$. It is worth mentioning that in the limit $\omega\rightarrow 0$, the particle density has a finite saturated value rather than an infrared divergence for the gapless case \cite{Antonin2012}. Also,  for large frequency $\omega \gg \kappa$, it is expected that the incoming mode from the past null infinity will
travel directly through the future horizon and  toward the future null infinity. Also, the modes from the Unruh states in the past horizon with such large frequencies will travel to the future null infinity. As such, $ e^{\frac{\omega}{2T_\text{hw}}}R_H^l\rightarrow 0,\, R_I^l\rightarrow 0$, and  $\mathcal{T}_H^l\rightarrow1$ render the expression of $n^{ul}$ having a tail of the exponential decay in frequency.

As compared with the $\varphi_{\omega,u_l}^{\text{out},l}$ mode, the other particle spectrum of emission inside the analog horizon is $\varphi_{\omega,v_l}^{\text{out},l}$ of the positive norm state where its particle density will have no contribution from the modes in the subsonic region because of the lack of mode mixing giving $R_I^l=0$, and it becomes
\begin{align}\label{nvl}
n^{vl}&=\langle U\vert (\hat{a}^{\text{out},l}_{\omega,v_l})^\dagger\,\hat{a}^{\text{out},l}_{\omega,v_l}\vert U\rangle\nonumber\\
&=\frac{1}{e^{\frac{\omega}{T_\text{hw}}}-1}\big\vert  e^{\frac{\omega}{2T_\text{hw}}}\mathcal{R}_H^{l\,\ast}+T_H^l\big\vert^2
\end{align}
with the coefficients in \eqref{Thl} and \eqref{mRhl} shown in Fig.~\ref{fig_nur} given respectively by the Unruh state in the past horizon in the $l$ and $r$ regions.
The particle densities satisfy  $n^{ur}+n^{vl}=n^{ul}$ for $\omega>\omega_r$ and $n^{vl}=n^{ul}$ for $\omega<\omega_r$, which can be verified by using the unitary relation $|R_H^l|^2-|\mathcal{T}_H^l|^2+|R_I^l|^2=-1$ accompanied with \eqref{conservation_hr} and \eqref{conservation_hl}.
The particle densities of  $n^{ul}$ and $n^{vl}$ share the same feature that they do not have a smooth change across $\omega_r$.

Here we come to study the mode correlator such as the correlator of the analog Hawking mode $\varphi_{\omega,u_r}^{\text{out},l}$ in the subsonic region and its partner $\varphi_{\omega,u_l}^{\text{out},l}$ in the supersonic region as well as the correlator of the $\varphi_{\omega,u_l}^{\text{out},l}$ mode and the $\varphi_{\omega,v_l}^{\text{out},l}$, both of which are in the supersonic regions.

The $ul \textendash ur$ correlator can be computed from
\begin{align}\label{curul}
	c_{ur,ul}=&\langle U\vert ^\text{out}\hat{a}^r_{\omega,u}\,^\text{out}\hat{a}^l_{\omega,u}\vert U\rangle\nonumber\\
	 =&R_I^{l\ast}R_I^r+\frac{e^{\frac{\omega}{2T_\text{hw}}}}{e^{\frac{\omega}{T_\text{hw}}}-1}\left(T_H^r\mathcal{T}_H^l+T_H^rR_H^{l\ast}e^{\frac{\omega}{2T_\text{hw}}}\right),
\end{align}
where the coefficients can be found in \eqref{thr}, \eqref{Rhl}, \eqref{RIr}-\eqref{RIl},  \eqref{mRhl}, and \eqref{mThl}.
Nevertheless, the $ul\textendash vl$ correlator can be obtained as
\begin{align}\label{cvlul}
	c_{vl,ul}=&\langle U\vert ^\text{out}\hat{a}^l_{\omega,v}\,^\text{out}\hat{a}^l_{\omega,u}\vert U\rangle\nonumber\\
	 =&R_I^{l\ast}T_H^l+\frac{1}{e^{\frac{\omega}{T_\text{hw}}}-1}\no\\
	 \times&\left[R_H^{l\ast}T_I^le^{\frac{\omega}{T_\text{hw}}}+(\mathcal{T}_H^lT_I^l+\mathcal{R}_H^{l\ast}R_H^{l\ast})e^{\frac{\omega}{2T_\text{hw}}}+ \mathcal{T}_H^{l}\mathcal{R}_H^{l\ast}\right],
\end{align}
where the coefficients can be found in \eqref{thr}, \eqref{Rhl}, \eqref{RIr}, \eqref{TIl}, \eqref{mRhl}, and \eqref{mThl}.

Both correlators as a function of frequency are shown in Fig. \ref{fig_ulur}. Because there is no particle production below the threshold frequency $\omega_r$ in  the $\varphi_{\omega,u_r}^{\text{out},r}$ mode of the analog Hawking mode, the correlator $c_{ur,ul}$ vanishes although the partner mode of the $\varphi_{\omega,u_l}^{\text{out},l}$ mode of the negative norm state dose have  particle production within this frequency range. For $\omega >\omega_r$, the correlator $c_{ur,ul}$ shows a peak around the threshold frequency. As for the correlator of $c_{vl,ul}$, apart from the large correlation in small frequency, there exists also a peak near the threshold frequency.
 Interestingly, as $\omega >\omega_r$, the magnitude of correlator $\vert c_{ur,ul}\vert$ for a gapped mode will be larger than gapless cases, especially around the threshold frequency.
 The behavior of the mode correlators will contribute to the density-density correlators to be done in our forthcoming work that can be measured experimentally \cite{Jeff2015,Jeff2019, Kolobov:2021wd}.

\section{summary and conclusion}\label{sec_conclusions}\label{ABH6}
We start from considering the condensates of cold atoms at zero temperature in the tunable binary BEC system  with  the Rabi transition between  atomic hyperfine states where the system can be represented by a coupled two-field model of gapless excitations and gapped excitations. For the general spatial-dependent coupling constant strengths,  the decoupling of two excitations under certain conditions of the condensate wave functions and the coupling constants is reviewed. In particular, we will solely focus on the dynamics of gapped excitations. The dispersion relation of the gapped modes involves the $k^2$ term in a  very long wavelength approximation that behaves relativistically.
The particular  spatial-dependent sound speed and flow velocity with the acoustic horizon in the elongated condensates are introduced so that the equations of the mode functions can be analytically treatable.  In addition, the horizon  generated from the transonic  flow is formed with experimentally accessible parameters.
As compared with the gapless excitations, there exists a threshold frequency $\omega_r$ in the subsonic region above which the  modes can propagate. The asymptotic states of the incoming and outgoing modes are identified where the scattering coefficients between them for various scattering processes can be  achieved. Accordingly, the Bogoliubov transformations of the creation and annihilation operators associated with the incoming and outgoing modes are derived.  Also,  the Unruh state
is introduced to be the appropriate state for the description of
gravitational collapse of the black hole.
The particle spectrum of the analogous Hawking modes in the exterior of the horizon of the subsonic region is computed, and is shown as a thermal one with temperature given by the analogous surface gravity $\kappa$, mainly due to the introduction of the Unruh state in the past horizon.
The associated graybody factor
significantly deviates from that of the gapless
cases near the threshold frequency, which vanishes as the
mode frequency is below  $\omega_r$.
In the interior region of the  horizon of the supersonic region, the spectrum of  the  particle production of the Hawking partner has the nonthermal feature. The correlators between the Hawking mode and its partner of relevance to the experimental observations show some peaks near the
threshold frequency $\omega_r$ resulting from  the gap energy term.  The behavior of the mode correlators will contribute to the density-density correlators  that can be measured experimentally to be carried out in our forthcoming work.

\begin{acknowledgements}
This work was supported in part by the
Ministry of Science and Technology, Taiwan, R.O.C.
\end{acknowledgements}

\end{document}